# Studies on the formation of formaldehyde during 2-ethylhexyl 4-(dimethylamino)benzoate demethylation in the presence of reactive oxygen and chlorine species

## Waldemar Studziński, Alicja Gackowska, Maciej Przybyłek & Jerzy Gaca

# Studies on the formation of formaldehyde during 2-ethylhexyl 4-(dimethylamino)benzoate demethylation in the presence of reactive oxygen and chlorine species

Waldemar Studziński[1] · Alicja Gackowska[1] · Maciej Przybyłek[2] · Jerzy Gaca[1]

**Abstract** In order to protect the skin from UV radiation, personal care products (PCPS) often contain chemical UV-filters. These compounds can enter the environment causing serious consequences on the water ecosystems. The aim of this study was to examine, the effect of different factors, such as UV light, the presence of NaOCl and $H_2O_2$ on the formaldehyde formation during popular UV filter, 2-ethylhexyl 4-(dimethylamino)benzoate (ODPABA) demethylation. The concentration of formaldehyde was determined by VIS spectrophotometry after derivatization. The reaction mixtures were qualitatively analyzed using GC/MS chromatography. The highest concentration of formaldehyde was observed in the case of ODPABA/$H_2O_2$/UV reaction mixture. In order to describe two types of demethylation mechanisms, namely, radical and ionic, the experimental results were enriched with Fukui function analysis and thermodynamic calculations. In the case of non-irradiated system containing ODPABA and NaOCl, demethylation reaction probably proceeds via ionic mechanism. As it was established, amino nitrogen atom in the ODPABA molecule is the most susceptible site for the HOCl electrophilic attack, which is the first step of ionic demethylation mechanism. In the case of irradiated mixtures, the reaction is probably radical in nature. The results of thermodynamic calculations showed that abstraction of the hydrogen from $N(CH_3)_2$ group is more probable than from 2-ethylhexyl moiety, which indicates higher susceptibility of $N(CH_3)_2$ to the oxidation.

## Introduction

In the recent 20 years, the annual consumption of pharmaceuticals and personal care products (PPCPs) has dramatically increased (Tong et al. 2011). Among PPCPs, sunscreen agents deserve particular attention. UV filters have been detected in wastewater, surface water (Poiger et al. 2004; Balmer et al. 2005; Ma et al. 2016), sewage sludge (Eljarrat et al. 2012; Zuloaga et al. 2012), river sediments (Amine et al. 2012; Kaiser et al. 2012), bathing waters and swimming pool waters (Vila et al. 2016; Ekowati et al. 2016), and even in drinking water (da Silva et al. 2015). The last example indicates the difficulty of UV filter elimination during waste water treatment. Nowadays, the major concern of UV filter contamination is their effect on the endocrine system of aquatic organisms (Krause et al. 2012; Kinnberg et al. 2015). Another important issue is the environmental fate of sunscreen agents. Recently, there has been a growing interest in the UV filter degradation research (Pattanaargson and Limphong 2001; Díaz-Cruz and Barceló 2009; Nakajima et al. 2009; Santos

🖂 Springer

et al. 2012; De Laurentiis et al. 2013; Santos et al. 2013; Gackowska et al. 2014; Hanson et al. 2015; Vione et al. 2015; Gackowska et al. 2016b; Gackowska et al. 2016a; Li et al. 2016; Tsoumachidou et al. 2016). Many of these studies included the effect of oxidizing agents on the sunscreen active ingredient degradation.

Advanced oxidation processes (AOPs) are efficient water treatment methods utilizing reactive oxygen species generation. Some examples of frequently used and studied AOPs are $TiO_2/UV$ (Hupka et al. 2006; Thiruvenkatachari et al. 2008), $H_2O_2$, $H_2O_2/UV$, $O_3$, $O_3/UV$ (Baus et al. 2007; Souza et al. 2016), $Fe^{2+}/H_2O_2$, $Fe^{3+}/H_2O_2$ (Gaca et al. 2005; Tong et al. 2011; Khankhasaeva et al. 2012), $Fe^{3+}/H_2O_2/UV$ (Kumar et al. 2008; Diagne et al. 2009; Li et al. 2012b; Topac and Alkan 2016; Tsoumachidou et al. 2016) and $Fe^{2+}/UV/S_2O_8^{2-}$ (Khan et al. 2013; Brienza et al. 2014; Xue et al. 2016). Unfortunately, in some cases, AOPs fail in formaldehyde elimination or even contributes to its generation (Can and Gurol 2003; Wert et al. 2007; Trenholm et al. 2008; Tripathi et al. 2011; Li et al. 2012a). However, when considering Fenton-like systems, the effectiveness of formaldehyde removal from its solutions is quite high, even 94% (Murphy et al. 1989; Kajitvichyanukul et al. 2006; Kowalik 2011; Guimarães et al. 2012; Méndez et al. 2015).

As it was reported (Emri et al. 2004), even very low concentrations of this aldehyde ($<10^{-4}$ M) causes DNA damage in human skin cells. The environmental occurrence of formaldehyde is caused by the anthropogenic and non-anthropogenic organic matter oxidation processes, by the release from resins and from other products, to which it is usually added as a preservative (Barker et al. 1996; Salthammer et al. 2010; Madureira et al. 2016; Ochs et al. 2016). One of the possible natural routes of formaldehyde entrance into the environment is through biochemical $O$- and $N$-demethylation being a part of metabolic conversions occurring in living organisms (Kalász 2003; Hagel and Facchini 2010; Farrow and Facchini 2013). However, there is also non-biochemical in nature processes of methyl group abstraction. For instance, formaldehyde can be formed as a result of amine demethylation in the presence of a disinfecting agent, HOCl (Mitch and Schreiber 2008; Kosaka et al. 2014). Furthermore, the formation of low-molecular-weight aldehydes including formaldehyde was postulated in the case of photo-induced radical dealkylation (Bozzi et al. 2002; Görner and Döpp 2003; Baciocchi et al. 2005). It is worth to mention that potential formaldehyde precursors with $N$-alkylated groups such as drugs (theophylline, caffeine, metamizole, phenazone, aminophenazone), dyes (methylene blue, methyl orange, crystal violet, malachite green), or quaternary ammonium surfactants are widely used in industry and households and therefore can enter the aquatic environment (Boethling 1984; Forgacs et al. 2004; Favier et al. 2015; Zhang et al. 2015). The main purpose of this paper is to evaluate whether degradation of a popular representative of this class, UV filter 2-ethylhexyl 4-(dimethylamino)benzoate, also known as octyl-dimethyl-p-aminobenzoic acid (ODPABA), can be a potential source of formaldehyde contamination. In order to get a better inside into the nature of ODPABA demethylation, the local reactivity analysis and thermodynamic calculations based on the density functional theory (DFT) were performed. Noteworthy, in the recent decade, quantum-chemical methods including thermodynamic calculations, reaction path modeling, reactivity analysis, and QSAR studies have been increasingly used in environmental studies (Cysewski et al. 2006; Kurtén et al. 2007; Blotevogel et al. 2010; Tröbs et al. 2011; Gaca et al. 2011; Elm et al. 2013; Gackowska et al. 2014; Turkay et al. 2015; Altarawneh and Dlugogorski 2015; Kurtén et al. 2015; Mamy et al. 2015; Xie et al. 2015; Myllys et al. 2016; Gackowska et al. 2016a; Shah and Hao 2016). Fukui function analysis is a robust and effective approach for evaluating the susceptibility of the individual atoms to the nucleophilic, electrophilic, and radical attack (Langenaeker et al. 1992; Pilepić and Uršić 2001; Özen et al. 2003; Martínez et al. 2009; De Witte et al. 2009; Rokhina and Suri 2012; Barr et al. 2012; Saha et al. 2012; Allison and Tong 2013; Altarawneh and Dlugogorski 2015). Previous studies showed that Fukui function can be successfully used in describing degradation and chlorination of a popular sunscreen agent 2-ethylhexyl 4-methoxycinnamate (Gackowska et al. 2014; Gackowska et al. 2016a). According to our best knowledge, there is no information in the literature about the ODPABA local reactivity properties and their potential consequences on the environmental fate. Therefore, the additional aim of this paper is to utilize Fukui function analysis to describe ODPABA degradation.

## Materials and methods

### Materials

All chemicals were purchased from commercial suppliers and used without purification. 2-ethylhexyl 4-(dimethylamino)benzoate (ODPABA, CAS: 21245-02-3) was obtained from Sigma-Aldrich (USA). Sodium hypochlorite NaOCl with a nominal free chlorine content of 100 g/L and $H_2O_2$ (30%) were obtained from POCh (Poland).

### Reaction conditions

The reaction mixtures were prepared by dissolving the reactants in 1000 ml of water according to the proportions given in Table 1. The effect of UV irradiation was examined with the use of photoreactor equipped with a Heraeus, TQ 150W

Table 1 The reaction conditions and substrate proportions used in this study

| Reagents | ODPABA [mM] | $H_2O_2$ [mM] | NaOCl [mM] | UV [W] | pH range |
|---|---|---|---|---|---|
| ODPABA/UV | 0.36 | 0 | 0 | 150 | 8.44–8.01 |
| ODPABA/NaOCl | 0.36 | 0 | 10 | – | 10.45–10.29 |
| ODPABA/NaOCl/UV | 0.36 | 0 | 10 | 150 | 10.30–8.11 |
| ODPABA/$H_2O_2$ | 0.36 | 10 | 0 | – | 8.31–8.05 |
| ODPABA/$H_2O_2$/UV | 0.36 | 10 | 0 | 150 | 8.42–6.82 |

medium pressure mercury lamp (200–600 nm), magnetic stirrer (200 rpm), and pH meter.

### Formaldehyde determination

Formaldehyde was determined in the reaction mixtures using Method 8110 Powder Pillows test kit and DR3900 Benchtop VIS Spectrophotometer provided by Hach, USA. This procedure of formaldehyde determination was designed for water samples by Hach company, based on the older colorimetric method used for air analysis (Matthews and Howell 1981). According to the procedure, the samples are derivatized with 3-methyl-2-benzothiazoline hydrazone (MBTH) using the equipment and chemicals provided in the test kit and then the resulting blue dye is determined through visible spectrophotometry ($\lambda_{max}$ = 630 nm).

### GC/MS measurements

After 180 min, the reaction mixture samples (50 mL) were extracted for 10 min by 1:1 n-hexane:ethyl acetate (10 mL). Then, so-prepared extracts were dried with anhydrous sodium sulfate. The ODPABA transformation products were detected with the use of 5890 HEWLETT PACKARD gas chromatographer equipped with a MS detector and the ZB-5MS column (0.25 mm × 30 m × 0.25 μm). The following chromatographic conditions were applied: sample volume 1 μL, helium carrier gas, injector temperature 250 °C, oven temperature program from 80 to 260 °C at 10 °C/min, from 260 to 300 °C at 5 °C/min.

### Quantum-chemical calculations

The geometry optimizations, frequencies, and thermochemical calculations were carried out at B3LYP/6-31+G(d,p) level (Krishnan et al. 1980; McLean and Chandler 1980; Clark et al. 1983; Frisch et al. 1984; Lee et al. 1988; Becke 1988; Miehlich et al. 1989; Becke 1993) with Gaussian03 software (Frisch et al. 2003). In the case of open shell reaction intermediates (radicals), unrestricted procedure (Čársky and Hubač 1991) was applied. In order to include the effect of the solvent on the molecular structure, polarized continuum model (PCM) was used (Miertuš et al. 1981; Miertuš and Tomasi 1982). All structures considered in this paper were checked for the absence of imaginary frequencies. Exemplary structural data and frequencies calculated for ODPABA are provided in Online Resource (Tables S1, S2, S3, and S4). Thermodynamic analysis was performed utilizing enthalpy values calculated automatically along with frequencies values according to the approach presented by Ochterski (2000).

Fukui function values (electrophilic $f^-$, nucleophilic, $f^+$ and radical $f^0$) were calculated according to previously reported method (Thanikaivelan et al. 2002; Gackowska et al. 2016a) using BLYP functional (Lee et al. 1988; Becke 1988; Miehlich et al. 1989) with DND basis set (version 3.5) (Delley 2006) and Hirshfeld charge population analysis (Hirshfeld 1977; Ritchie 1985; Ritchie and Bachrach 1987). All these computations were performed within DMOL$^3$ (Delley 1990; Delley 1996; Delley 2000) module of Accelrys Material Studio 7.0 (Accelrys Materials Studio 7 2014). In this study, Fukui function values were calculated according to the Yang and Mortier procedure (Yang and Mortier 1986) using the following Eqs. (1–3):

$$f^+ = Q(n+1) - Q(n) \quad (1)$$

$$f^- = Q(n) - Q(n-1) \quad (2)$$

$$f^0 = \frac{Q(n+1) - Q(n-1)}{2} \quad (3)$$

where $Q$ denotes the Hirshfeld charge and $n$ the number of electrons in the molecule.

### Results and discussion

At the first stage of the study, the effect of popular water treatment and disinfection agents $H_2O_2$, NaOCl, and UV irradiation on the formaldehyde formation was examined. The relationships between formaldehyde concentration increase and reaction time is presented on Fig. 1. As we found, the highest concentration of formaldehyde was reached in the case of irradiated samples. It is worth to mention that according to some studies (Mopper and Stahovec 1986; Kieber et al. 1990; Zhou and Mopper 1997; Reader and Miller 2011), there is a relationship between the presence of dissolved organic matter capable of absorbing UV light and formaldehyde contamination.

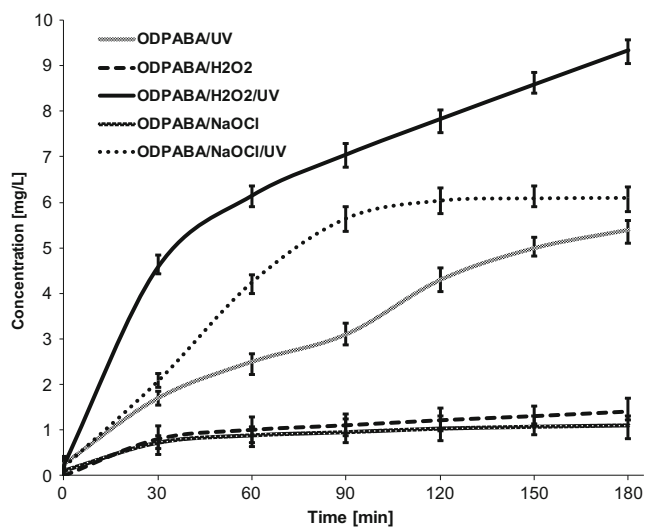

**Fig. 1** The effects of different agents on the formation of formaldehyde during ODPABA demethylation

Interestingly, the concentration of formaldehyde in the case of ODPABA/UV system is higher than in the case of ODPABA/NaOCl and ODPABA/$H_2O_2$. This shows that photo-induced demethylation occurs readily even without oxidizing agent addition. In general, two types of photodegradation reactions can be distinguished, namely, direct and indirect. According to the indirect photolysis mechanism, demethylation can be caused by the reactive species formation in the presence of UV filter acting as a photosensitizer. It has been demonstrated that the occurrence of chromophoric dissolved organic matter (CDOM) contributes to the formation of various transients such as carbonate radical ($CO_3^{-•}$), hydroxyl radical ($^•OH$), singlet oxygen ($^1O_2$), and excited triplet state $^3CDOM^*$ (Schwarzenbach et al. 2005; Vione et al. 2010; Kelly and Arnold 2012; Vione et al. 2014; De Laurentiis et al. 2014; Vione et al. 2015; McNeill and Canonica 2016; Vione 2016). Noteworthy, the presence of ODPABA and $p$-aminobenzoic acid (PABA) significantly enhance singlet oxygen ($^1O_2$) generation in UV-irradiated water (Allen and Gossett 1996). In the case of ODPABA/NaOCl/UV, system demethylation is probably induced by the radical attack of reactive chlorine and oxygen species such as $^•OH$, $Cl^•$, $^•OCl$, $^•O^-$, $Cl_2^{-•}$, and $ClOH^{-•}$ formed in the presence of chlorinating agent (Feng et al. 2007; Chan et al. 2012; Fang et al. 2014; Vione et al. 2014). Since ODPABA absorption bands are overlapped with emission spectrum bands of the lamp used in this study (supplementary Fig. S1), direct photodegradation is another potential pathway, which can occur under irradiation conditions. Noteworthy, both direct and indirect mechanisms were found to be a possible explanation of popular sunscreen agent, 2-ethylhexyl 4-methoxycinnamate degradation in surface waters (Vione et al. 2015); however, direct photodegradation was found to be the main route.

As it was reported in previous studies (Sakkas et al. 2003; Nakajima et al. 2009; Calza et al. 2016; Gackowska et al. 2016b), ODPABA demethylation products were formed in the case of ODPABA/UV, ODPABA/NaOCl, and ODPABA/NaOCl/UV systems. The GC/MS analysis presented in this paper confirmed this observation, since identified mass spectra of demethylated degradation ODPABA products are in accordance with literature data (Sakkas et al. 2003). Compounds containing $NH_2$ and $NH(CH_3)$ groups were also detected in the case of ODPABA/$H_2O_2$ and ODPABA/$H_2O_2$/UV systems, which have not been studied before. Retention times of detected compounds in the reaction mixtures are given in Table 2. Exemplary mass spectra are provided in Online Resource.

As it was reported (MacManus-Spencer et al. 2011; Gackowska et al. 2014; Gackowska et al. 2016b; Gackowska et al. 2016a), 2-ethylhexyl esters easily undergoes decomposition resulting 2-ethylhexanol. The same observation can be done for the reaction mixtures considered in this study. Since mass spectra of 2-ethylhexyl can be found in the National Institute of Standards and Technology (NIST) database (https://www.nist.gov/), it can be easily identified.

According to fragmentation pathways of ODPABA derivatives presented by Nakajima et al. (2009), a characteristic McLafferty rearrangement can be observed on the mass spectra. As a result of this reaction, ODPABA molecular ion decomposes into neutral 3-methyleneheptane ($C_8H_{16}$) and 4-($N,N$-dimethyl)aminobenzoic acid cation radical (m/z = 166) which corresponds to the most intense peak (Fig. S4 in the Online Resource). Analogical fragmentation behavior is observed in case of demethylated ODPABA derivatives (supplementary Figs. S2, S5, S6, and S7), since mass spectra recorded for those compounds are characterized by the low molecular peaks and the loss of 112 atomic mass units due to the 3-methyleneheptane molecule elimination.

The presence of dichlorinated compounds in ODPABA/NaOCl and ODPABA/NaOCl/UV reaction mixtures can be evidenced by the characteristic chlorine isotope signature. Since chlorine occurs in the form of two major isotopes, namely, $^{35}Cl$ (c.a. 76%) and $^{37}Cl$ (c.a. 24%), MS spectra of chlorinated compounds are characterized by the specific patterns. Depending on the number of chlorine atoms in the molecule, a different isotope signature is observed. The presence of two atoms causes the appearance of three m/z peaks M (high intensity), M+2 (lower intensity), and M+4 (the lowest intensity), due to the three isotope combinations ($^{35}Cl/^{35}Cl$, $^{35}Cl/^{37}Cl$, and $^{37}Cl/^{37}Cl$). Hence, in the case of dichlorinated 2-ethylhexyl 4-(methylamino)benzoate, the most intense peak (m/z = 219) is near to the MS signals at m/z = 221 and m/z = 223 (supplementary Fig. S6). A similar pattern can be observed for dichlorinated 2-ethylhexyl 4-aminobenzoate (Fig. S7).

Although the use of hydrogen peroxide is preferable from the viewpoint of avoiding unwanted chlorinated compounds, it contributes to the formation of significant quantities of formaldehyde (Fig. 1). This is caused by the rapid photodecomposition of $H_2O_2$ yielding hydroxyl radicals ($OH^\bullet$). According to the studies on photo-induced dealkylation including demethylation of p-substituted N,N-dimethylaniline derivatives (Bozzi et al. 2002; Görner and Döpp 2003; Baciocchi et al. 2005; Podsiadły et al. 2007), the reaction proceeds through one-electron oxidation followed by the deprotonation of cation radical resulting in $R\text{-}Ph\text{-}N(CH_3)_2CH_2^\bullet$ radical formation (Fig. 2a). This mechanism seems to be a highly probable explanation of N,N-dialkylated aromatic amine photodegradation, since the reaction intermediates namely cation radicals and radicals were observed using spectroscopic methods (Görner and Döpp 2002; Zielonka et al. 2004; Podsiadły et al. 2007). On the other hand, when ODPABA degradation is carried out in the presence of NaOCl, the reaction probably proceeds via ionic mechanism. As it was reported in the literature (Ellis and Soper 1954; Mitch and Schreiber 2008; Kosaka et al. 2014), amine demethylation proceeds via the following steps: electrophilic substitution of HOCl, elimination of HCl, water addition to the N=$CH_2$ bond, and finally formaldehyde elimination (Fig. 2b).

**Table 2** Retention times and selected MS data of detected compounds

| Reaction system | Detected compound | Linear formula | Retention time (min.) |
|---|---|---|---|
| ODPABA/UV | 2-ethylhexyl 4-aminobenzoate | $H_2NC_6H_4COOCH_2CH(C_2H_5)(CH_2)_3CH_3$ | 18.02 |
| | 2-ethyl-1-hexanol | $CH_3(CH_2)_3(C_2H_5)CHCH_2OH$ | 4.17 |
| ODPABA/$H_2O_2$ | ODPABA | $(CH_3)_2NC_6H_4COOCH_2CH(C_2H_5)(CH_2)_3CH_3$ | 19.44 |
| | 2-ethylhexyl 4-(methylamino)benzoate | $(CH_3)HNC_6H_4COOCH_2CH(C_2H_5)(CH_2)_3CH_3$ | 19.11 |
| | 2-ethyl-1-hexanol | $CH_3(CH_2)_3(C_2H_5)CHCH_2OH$ | 4.15 |
| ODPABA/$H_2O_2$/UV | 2-ethylhexyl 4-(methylamino)benzoate | $(CH_3)HNC_6H_4COOCH_2CH(C_2H_5)(CH_2)_3CH_3$ | 19.24 |
| | 2-ethylhexyl 4-aminobenzoate | $H_2NC_6H_4COOCH_2CH(C_2H_5)(CH_2)_3CH_3$ | 18.16 |
| | 2-ethyl-1-hexanol | $CH_3(CH_2)_3(C_2H_5)CHCH_2OH$ | 4.16 |
| ODPABA/NaOCl | dichlorinated 2-ethylhexyl 4-(methylamino)benzoate | $(CH_3)HNC_6H_4Cl_2COOCH2CH(C_2H_5)(CH_2)_3CH_3$ | 20.38 |
| | dichlorinated 2-ethylhexyl 4-aminobenzoate | $H_2NC_6H_4Cl_2COOCH2CH(C_2H_5)(CH_2)_3CH_3$ | 19.79 |
| | ODPABA | $(CH_3)_2NC_6H_4COOCH_2CH(C_2H_5)(CH_2)_3CH_3$ | 19.50 |
| | 2-ethyl-1-hexanol | $CH_3(CH_2)_3(C_2H_5)CHCH_2OH$ | 4.16 |
| ODPABA/NaOCl/UV | dichlorinated 2-ethylhexyl 4-(methylamino)benzoate | $(CH_3)HNC_6H_4Cl_2COOCH2CH(C_2H_5)(CH_2)_3CH_3$ | 20.27 |
| | 2-ethyl-1-hexanol | $CH_3(CH_2)_3(C_2H_5)CHCH_2OH$ | 4.15 |

The local reactivity of particular atoms in the molecule can be quantitatively evaluated using conceptual density functional theory. The higher the value of Fukui function of considered atom, the greater its reactivity. At the next stage of this study, in order to explain the nature of possible ODPABA demethylation mechanisms, quantum-chemical calculations including Fukui function analysis were performed. Since ODPABA can undergo hydrolysis in the environment, the local reactivity analysis was extended for ODPABA degradation product, p-(dimethylamino)benzoic acid (DMABA) and its zwitterionic tautomer (DMABA-ZW). The Fukui function values calculated for ODPABA, DMABA, and DMABA-ZW are summarized in Table 3. Optimized structures along with atom numbering scheme are given in Fig. 3.

The $f$ index values calculated for ODPABA and DMABA are the greatest in the case of amino nitrogen atom. Therefore,

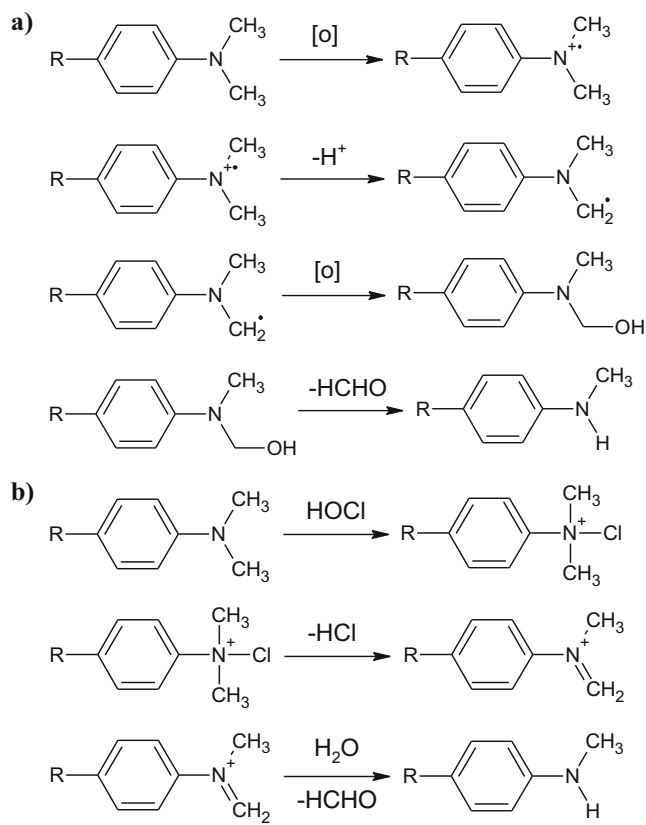

**Fig. 2** Radical (**a**) and ionic (**b**) demethylation mechanism of amines shown on the example of p-substituted N,N-dimethylaniline

**Table 3** Radical $f^0$, nucleophilic $f^+$, and electrophilic $f^-$ Fukui function values calculated for hydrogen atoms in ODPABA, DMABA, and its zwitterionic tautomer DMABA-ZW (atom numbering according to Fig. 3)

| Atom | ODPABA | | | Atom | DMABA | | | Atom | DMABA (ZW) | | |
|---|---|---|---|---|---|---|---|---|---|---|---|
|  | $f^+$ | $f^-$ | $f^0$ |  | $f^+$ | $f^-$ | $f^0$ |  | $f^+$ | $f^-$ | $f^0$ |
| H1 | 0.022 | 0.028 | 0.025 | H1 | 0.023 | 0.028 | 0.026 | H1 | 0.036 | 0.018 | 0.027 |
| H2 | 0.023 | 0.040 | 0.032 | H2 | 0.022 | 0.036 | 0.029 | H2 | 0.023 | 0.006 | 0.014 |
| H3 | 0.019 | 0.034 | 0.026 | H3 | 0.022 | 0.036 | 0.029 | H3 | 0.024 | 0.007 | 0.016 |
| H4 | 0.022 | 0.028 | 0.025 | H4 | 0.023 | 0.028 | 0.026 | H4 | 0.052 | 0.018 | 0.035 |
| H5 | 0.023 | 0.040 | 0.032 | H5 | 0.022 | 0.036 | 0.029 | H5 | 0.022 | 0.006 | 0.014 |
| H6 | 0.018 | 0.033 | 0.026 | H6 | 0.022 | 0.036 | 0.029 | H6 | 0.019 | 0.007 | 0.013 |
| H7 | 0.028 | 0.031 | 0.029 | H7 | 0.029 | 0.032 | 0.031 | H7 | 0.040 | 0.030 | 0.035 |
| H8 | 0.032 | 0.029 | 0.031 | H8 | 0.034 | 0.030 | 0.032 | H8 | 0.030 | 0.026 | 0.028 |
| H9 | 0.028 | 0.031 | 0.029 | H9 | 0.029 | 0.033 | 0.031 | H9 | 0.034 | 0.028 | 0.031 |
| H10 | 0.033 | 0.030 | 0.032 | H10 | 0.034 | 0.031 | 0.032 | H10 | 0.029 | 0.026 | 0.027 |
| H11 | 0.014 | 0.010 | 0.012 | H11 | 0.040 | 0.029 | 0.034 | H11 | 0.109 | 0.015 | 0.062 |
| H12 | 0.014 | 0.010 | 0.012 | C12 | 0.061 | 0.041 | 0.051 | C12 | 0.048 | 0.023 | 0.035 |
| H13 | 0.003 | 0.001 | 0.002 | C13 | 0.042 | 0.064 | 0.053 | C13 | 0.050 | 0.037 | 0.044 |
| H14 | −0.007 | −0.008 | −0.008 | C14 | 0.074 | 0.036 | 0.055 | C14 | 0.056 | 0.037 | 0.046 |
| H15 | 0.005 | 0.004 | 0.004 | C15 | 0.042 | 0.066 | 0.054 | C15 | 0.061 | 0.037 | 0.049 |
| H16 | 0.004 | 0.002 | 0.003 | C16 | 0.063 | 0.042 | 0.053 | C16 | 0.042 | 0.022 | 0.032 |
| H17 | 0.008 | 0.005 | 0.007 | C17 | 0.045 | 0.087 | 0.066 | C17 | 0.076 | −0.008 | 0.034 |
| H18 | 0.009 | 0.007 | 0.008 | N18 | 0.044 | 0.109 | 0.077 | N18 | 0.033 | 0.006 | 0.020 |
| H19 | 0.001 | 0.000 | 0.000 | C19 | 0.017 | 0.028 | 0.023 | C19 | 0.028 | 0.010 | 0.019 |
| H20 | 0.003 | 0.002 | 0.003 | C20 | 0.017 | 0.028 | 0.023 | C20 | 0.038 | 0.010 | 0.024 |
| H21 | 0.007 | 0.006 | 0.006 | C21 | 0.104 | 0.031 | 0.068 | C21 | 0.022 | 0.093 | 0.058 |
| H22 | 0.003 | 0.002 | 0.003 | O22 | 0.125 | 0.077 | 0.101 | O22 | 0.066 | 0.274 | 0.170 |
| H23 | 0.002 | 0.001 | 0.002 | O23 | 0.067 | 0.034 | 0.050 | O23 | 0.064 | 0.274 | 0.169 |
| H24 | 0.002 | 0.002 | 0.002 |  |  |  |  |  |  |  |  |
| H25 | 0.004 | 0.003 | 0.003 |  |  |  |  |  |  |  |  |
| H26 | 0.007 | 0.006 | 0.006 |  |  |  |  |  |  |  |  |
| H27 | 0.003 | 0.002 | 0.003 |  |  |  |  |  |  |  |  |
| C28 | 0.097 | 0.024 | 0.060 |  |  |  |  |  |  |  |  |
| C29 | 0.072 | 0.034 | 0.053 |  |  |  |  |  |  |  |  |
| C30 | 0.040 | 0.063 | 0.052 |  |  |  |  |  |  |  |  |
| C31 | 0.062 | 0.042 | 0.052 |  |  |  |  |  |  |  |  |
| C32 | 0.048 | 0.086 | 0.067 |  |  |  |  |  |  |  |  |
| C33 | 0.058 | 0.040 | 0.049 |  |  |  |  |  |  |  |  |
| C34 | 0.040 | 0.061 | 0.051 |  |  |  |  |  |  |  |  |
| N35 | 0.041 | 0.112 | 0.077 |  |  |  |  |  |  |  |  |
| C36 | 0.016 | 0.028 | 0.022 |  |  |  |  |  |  |  |  |
| O37 | 0.044 | 0.019 | 0.031 |  |  |  |  |  |  |  |  |
| O38 | 0.114 | 0.062 | 0.088 |  |  |  |  |  |  |  |  |
| C39 | 0.001 | 0.000 | 0.000 |  |  |  |  |  |  |  |  |
| C40 | 0.003 | 0.002 | 0.003 |  |  |  |  |  |  |  |  |
| C41 | −0.001 | −0.001 | −0.001 |  |  |  |  |  |  |  |  |
| C42 | 0.002 | 0.001 | 0.001 |  |  |  |  |  |  |  |  |
| C43 | 0.009 | 0.007 | 0.008 |  |  |  |  |  |  |  |  |
| C44 | 0.002 | 0.002 | 0.002 |  |  |  |  |  |  |  |  |
| C45 | 0.005 | 0.004 | 0.004 |  |  |  |  |  |  |  |  |
| C46 | 0.003 | 0.003 | 0.003 |  |  |  |  |  |  |  |  |
| C47 | 0.016 | 0.028 | 0.022 |  |  |  |  |  |  |  |  |

one may consider that ODPABA and DMABA readily undergo reactions involving this reactive site. This is not trivial observation since, in the case of aromatic amines, the amino group is conjugated with phenyl ring, which causes reduction of the negative charge density on the nitrogen atom. This effect is even more pronounced in the case of push-pull systems, i.e., when there is an electron accepting substituent attached in *para* position (as it is in the case of ODPABA and DMABA). Obviously, low negative charge density implies a low susceptibility of the atom to the electrophilic attack. However, in the case of ODPABA and DMBA, the nitrogen atom is even more suitable for electrophilic substitution than *orto*- (C30, C34) and *para*- (C32) positions in phenyl ring. Therefore, *N*-chlorination seems to be more favored than electrophilic substitution on the benzene ring. However, these results do not exclude the possibility of phenyl ring chlorination. Although in general aromatic amines are more prone for *C*-chlorination than for *N*-chlorination, there are some exceptions including drugs with *N*-alkylated moieties, e.g., fluoroquinolones (Weinberg et al. 2007). The resulting *N*-chlorinated aromatic amines can rearrange to the *C*-chlorinated compounds or decompose. *C*-chlorination is of course a competitive reaction to the HCl elimination step of demethylation mechanism (Fig. 2b). Noteworthy, as it was reported in the previous works (Sakkas et al. 2003; Gackowska et al. 2016b), both chlorinated and demethylated compounds were formed in ODPABA/NaOCl reaction mixture, which is also consistent with the results presented in this study (Table 2). In the case of zwitterionic form of ODPABA hydrolysis product, DMABA-ZW, the nitrogen atom is not susceptible for the electrophilic attack due to the attached H11 proton (Fig. 3c). Noteworthy, the $pK_{a1}$ and $pK_{a2}$ values of DMABA are 6.03 and 11.49, respectively (Haynes et al. 2014), and so, the isoelectric point pI is 8.76. Therefore, in the case of ODPABA/NaOCl system where the pH ranges from 10.45 to 10.29 (Table 1), the most dominant form would be neutral DMABA.

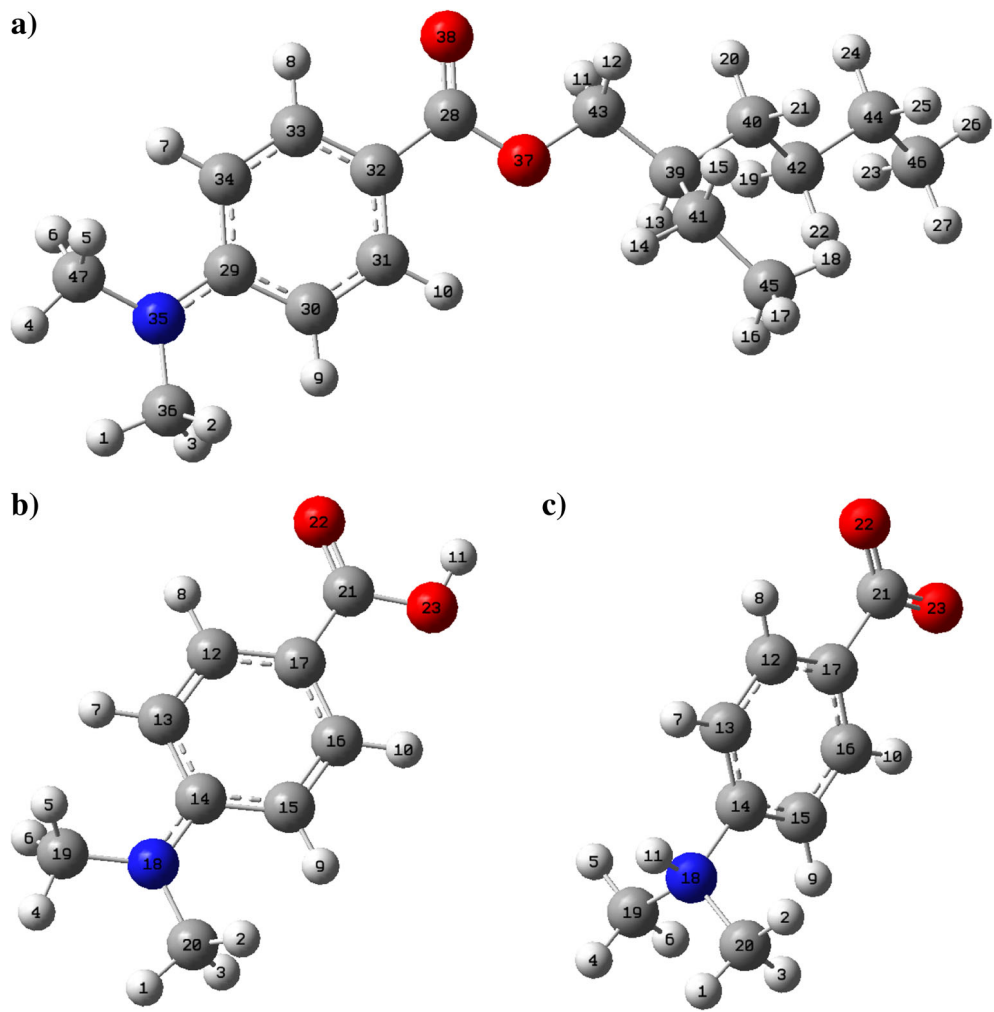

**Fig. 3** Visual representation of optimized molecular structures of ODPABA (**a**), DMABA (**b**), and its zwitterionic tautomer DMABA-ZW (**c**) along with atom numbering scheme

As it was mentioned, the R-Ph-N(CH$_3$)$_2$CH$_2^\bullet$ radical is one of the photo-induced demethylation intermediates (Fig. 2a). According to a different mechanism proposed for Michler ketone demethylation (Lu et al. 2009), the initial step of the reaction is radical attack to the hydrogen atom from the N(CH$_3$)$_2$ group. It is worth to note that the thermodynamic stability of reactive organic species like radicals and carbocations is an important factor determining which reaction pathway is more kinetically favored. This is so because, according to the Hammond rule (Hammond 1955), transition states are energetically similar to the reaction intermediates. Hence, these reaction paths are more preferred which involve low energy intermediates. The radical formed via hydrogen atom abstraction from the N(CH$_3$)$_2$ group is probably highly stabilized by the π-electron delocalization. Selected resonance structures illustrating unpaired electron delocalization in R-Ph-N(CH$_3$)$_2$CH$_2^\bullet$ radical are shown on supplementary Fig. S7 (Online Resource). In order to evaluate the stability of radicals formed via hydrogen atom abstraction from ODBABA molecule, quantum-chemical thermodynamic calculations of the hypothetical reactions with hydroxyl radical were performed (Fig. 4). This analysis showed that OH$^\bullet$ attack to the methyl group attached to nitrogen atom is the most thermodynamically favored (abstraction of H1-H6 atoms). Obviously, formation of phenyl radical is the least probable (H7-H10 abstraction), since it is well known that the radical stability decreases in the order tertiary > secondary > primary > phenyl. As one can see from Fig. 4, the enthalpy change of OH$^\bullet$ attack on the H13 atom is highly negative in comparison to other hydrogen atoms in 2-ethylhexyl moiety. The significant stability of formed in this reaction tertiary radical can be explained by the hyperconjugation effect. Nevertheless, according to the Fukui function, analysis OH$^-$ attack on 2-ethylhexyl moiety is highly unfavorable. As it can be inferred from Table 3, $f^0$ values of the 2-ethylhehyl group hydrogen atoms are significantly smaller than in other cases. Therefore, oxidation of methyl groups attached to amino nitrogen is the more preferable pathway of formaldehyde formation than oxidation of methyl groups in 2-ethylhexyl moiety.

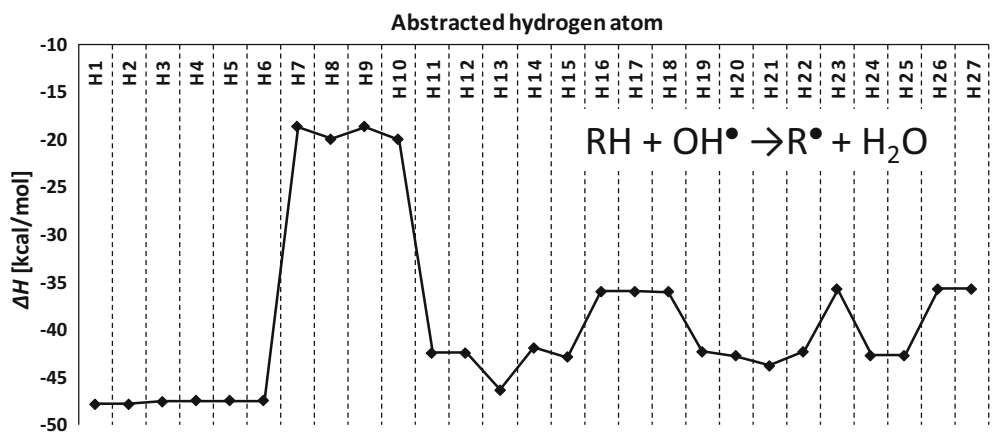

**Fig. 4** Enthalpy changes of the hydrogen atom abstraction reaction from ODPABA by hydroxyl radical (atom numbering according to Fig. 3a)

Although the above local reactivity and thermodynamic analysis was found to be consistent with experimental results, it should be taken into account that calculated values are strongly dependent on the computation level. In this study, a low-computational-cost method, namely, B3LYP/6-31+(d,p), was used. Due to its efficiency, B3LYP is probably the most extensively used functional including UV filter modeling (Alves et al. 2011; Corrêa et al. 2012; Ferreira et al. 2014; Miranda et al. 2014; Gackowska et al. 2014; Garcia et al. 2015; Gackowska et al. 2016a). It is worth mentioning that structures optimized using B3LYP functional and double zeta basis sets were successfully used for UV absorption property prediction of 2-ethylhexyl 4-methoxycinnamate (Alves et al. 2011; Miranda et al. 2014) and benzophenone sunscreen agents (Corrêa et al. 2012). Many studies demonstrated that thermodynamic parameters calculated using B3LYP functional and double zeta basis sets were in good accordance with experimental results (Muñoz-Muñiz and Juaristi 2002; Guner et al. 2003; Li et al. 2003; Ling

Qiu et al. 2006; Chirico et al. 2016). Nevertheless, the B3LYP approach does not include dispersion effects that might be important in the case of the molecules stabilized by the intramolecular interactions (Seebach et al. 2010; Steinmann et al. 2010; DiLabio et al. 2013).

## Conclusions

Since formaldehyde has been recognized as a toxic and carcinogenic compound, it has been numerous attempts to determine the anthropogenic and non-anthropogenic sources of its release to the environment. In this study, a popular UV filter ODPABA degradation in the presence of water treatment and disinfection agents (UV irradiation, NaOCl, $H_2O_2$) was examined. As it was established, the highest concentration of formaldehyde was achieved in the case of irradiated reaction mixtures. This is understandable, since *N*-alkylated aniline derivatives can undergo dealkylation under radical reaction conditions. On the other hand, ODPABA demethylation in the presence of water disinfecting-agent, NaOCl, probably proceeds via ionic mechanism.

Since the environmental fate of chemical compounds is closely related to their reactivity, quantum-chemical calculations can be used as a powerful tool in predicting and describing degradation pathways. It is worth to mention that this approach was utilized in the previous studies dealing with degradation of 2-ethylhexyl methoxy cinnamate (Gackowska et al. 2014; Gackowska et al. 2016a). According to presented herein, Fukui function values analysis, amino nitrogen atom is the most suitable for electrophilic substitution reaction, which is the initial step of ionic mechanism (Ellis and Soper 1954; Mitch and Schreiber 2008; Kosaka et al. 2014). Thermodynamic calculations showed that abstraction of the hydrogen atom from the $N(CH_3)_2$ group during radical demethylation reaction is more preferable than from 2-ethylhexyl moiety.

Experimental and theoretical studies on the emerging contamination degradation in the presence of various environmentally relevant agents are helpful in determining which chemical compounds may be formed during real processes taking place in nature and during wastewater treatment. Therefore, carrying out of such research may contribute to better monitoring of toxic compounds in the environment. The presented results in this paper indicate that the presence of ODPABA in water can cause a significant formaldehyde contamination. Therefore, also, other methyl group-containing compounds should be tested for the ability of formaldehyde formation when assessing the environmental risk. Since photo-induced ODPABA demethylation occurs readily even without oxidizing and chlorinating agent addition, there is a need to examine the concentration of formaldehyde in bathing and swimming pool waters. Moreover, formaldehyde even in very small amounts can cause DNA damage in the human skin cells (Emri et al. 2004). This is important in the context of possible mutagenic action affected by the release of this compound on the skin from ODPABA containing-cosmetics under the influence of UV radiation.

**Acknowledgments** We thank the Academic Computer Center in Gdańsk for providing its facilities to perform calculations presented in this paper.